\documentclass{iopart}
\usepackage{iopams}
\usepackage{graphicx}
\usepackage{times}

\newcommand{\bsy}[1]{\boldsymbol{#1}}
\newcommand{\dagg}{^{\dagger}}
 \newcommand{\degr}{^{\circ}}
\newcommand{\Wac}{W_{\mathrm{ac}}}

\begin{document}

\title{Interfaces Within Graphene Nanoribbons}

\author{J Wurm$^{1,2}$, M Wimmer$^{1}$, \.{I} Adagideli$^{1,3}$, K Richter$^{1}$ and \\H U Baranger$^{2}$}
\address{$^1$Institut f\"ur Theoretische Physik, Universit\"at Regensburg, D-93040 Regensburg, Germany}
\address{$^2$Department of Physics, Duke University,
Durham, North Carolina 27708-0305, U.S.A.}
\address{$^3$Faculty of Engineering and Natural Sciences, Sabanc\i~University, Orhanl\i~-- Tuzla, 34956, Turkey}
\eads{\mailto{Juergen.Wurm@physik.uni-regensburg.de}, \mailto{harold.baranger@duke.edu}}

\begin{abstract}
We study the conductance through two types of graphene nanostructures:
nanoribbon junctions in which the width changes from wide to narrow, and curved
nanoribbons. In the wide-narrow structures, substantial reflection occurs from
the wide-narrow interface, in contrast to the behavior of the much studied
electron gas waveguides. In the curved nanoribbons, the conductance is very
sensitive to details such as whether regions of a semiconducting armchair
nanoribbon are included in the curved structure -- such regions strongly
suppress the conductance. Surprisingly, this suppression is not due to the band
gap of the semiconducting nanoribbon, but is linked to the valley degree of
freedom.  Though we study these effects in the simplest contexts, they can be
expected to occur for more complicated structures, and we show results for rings
as well. We conclude that experience from electron gas waveguides does
\textit{not} carry over to graphene nanostructures. The interior interfaces
causing extra scattering result from the extra effective degrees of freedom of
the graphene structure, namely the valley and sublattice pseudospins.
\end{abstract}

\vspace*{-0.2in}
%

\section{Introduction}

There has been tremendous interest recently in investigating
carbon-based 
nanoelectronics, first with carbon nanotubes
\cite{Saito2003, Charlier2007, Avouris2009} and more recently with graphene \cite{CastroNeto2009}. 
In that
context, researchers have intensively studied graphene ``nanoribbons''
-- infinite, straight strips of graphene of constant width -- both theoretically
\cite{Nakada1996, Fujita1996,Wakabayashi2001, Ezawa2006, Brey2006a, Peres2006a, Munoz2006, Rycerz2007,Areshkin2007a,Xu2007,
 Cresti2007, Fernandez-Rossier2007, Zheng2007, Akhmerov2008,Rycerz2008,Katsnelson2008,Iyengar2008, Wimmer2008a, Mucciolo2009}
and experimentally
\cite{Chen2007, Han2007, Li2008, Wang2008, Datta2008, Tapaszto2008, Stampfer2009, Jiao2009}. 
Most of the theoretical effort has been focused on nanoribbons of essentially constant width.
However, more functionality, beyond that of a mere wire, might be gained if one considers more general and realistic nanoribbons 
in which the width of the ribbon changes, it curves, or particular junctions of nanoribbons are formed. 

On a more fundamental level, the continuing great interest in the effect of
reduced dimensionality, such as electron-electron interactions in
reduced dimensions, provides motivation for studying
quasi-one-dimensional systems. Graphene's unusual dispersion (massless
Dirac fermions) and reduced density of states at the Fermi energy, for
instance, suggest potential for novel effects. Of course, one should
first understand the non-interacting system before turning to
interactions.

Graphene nanoribbons are closely analogous to electron waveguides
patterned out of two dimensional electron gas (2DEG), usually in GaAs or 
other semiconductor systems
\cite{vanWees1988, Wharam1988, Timp1988, Takagaki1988, Szafer1989,  Sols1989, Baranger1990, Goodnick1991, Bagwell1993, Lent1994, Wang2007}. 
However, there is an important difference in
how the confinement is achieved. While in 2DEG waveguides the
electrons are trapped in the transverse direction of the waveguide by
applying a potential by means of local gate electrodes, graphene
nanoribbons are directly cut out of a larger graphene flake. This
gives rise to different types of boundaries, depending on the
direction in which the nanoribbons are cut with respect to the
graphene lattice. If the longitudinal direction of the nanoribbon is
along the direction of nearest neighbor carbon bonds, the resulting
boundary is of ``armchair'' type, while cutting at $30\degr$ with respect 
to the nearest neighbor
carbon bonds results in a ``zigzag'' boundary (see figure \ref{fig:figure1}). 
It has been shown that the low energy properties of nanoribbons with
boundaries other than these two are 
equivalent to those of zigzag nanoribbons \cite{Akhmerov2008a}.
On the experimental side, there has
been 
recent progress in controlling the edges of graphene samples
\cite{Jia2009, Girit2009}, which is essential to enable physicists to
probe the influence of edge details on transport properties.

This paper is organized as follows: First we study one of the most
simple systems beyond a straight nanoribbon with constant width,
namely wide-narrow junctions, by which we mean
two semi-infinite nanoribbons
attached together to form a step.
We calculate the
conductance of such ribbons by numerically solving the tight binding model,
and also obtain analytical results for the case of armchair boundaries.
In the second part we investigate numerically the
conductance of curved wires cut out of graphene. In this case the
width of the nanoribbon is approximately constant, but the longitudinal direction
with respect to the underlying graphene lattice and hence the
transverse boundary conditions change locally.  In contrast to systems
with sharp kinks and abrupt changes in the direction, which have been
investigated in earlier work 
\cite{Wakabayashi2001,Areshkin2007a,Xu2007,Rycerz2008,Katsnelson2008,Iyengar2008,Wurm2009}, 
we focus here on smooth bends. 

In both cases we find
remarkable deviations from the conductance of 2DEG waveguides that
are clear signatures of the sublattice and valley degrees of freedom in
the effective 2D
Dirac Hamiltionian describing graphene's low energy excitations,
\begin{equation}
H = v_F
\left(
\begin{array}{cc}
 \sigma_x p_x+\sigma_y p_y & 0 \\
0 & -\sigma_x p_x + \sigma_y p_y
\end{array}
\right)\;.
\label{eq:DiracHam}
\end{equation}
Here the matrix structure is in valley space, $\sigma_{x/y}$
are Pauli matrices in pseudo- or sublattice- spin space, $p_{x/y}$
are the momentum operators, and $v_F\approx 10^6\,$m/s is the Fermi
velocity.
Alternatively, from a strictly lattice point of view, the deviations that we see are
caused by the basis inherent in graphene's hexagonal lattice. 

For our numerical work, we use a nearest-neighbor tight binding model
taking into account the $2p_z$-orbitals of the carbon atoms 
\cite{CastroNeto2009, Wallace1947} and solve
the transport problem using an adaptive recursive Green function
method \cite{Wimmer2008} to obtain the conductance $G$. Throughout the
paper, lengths are given in units of the graphene lattice constant $a$
which is $\sqrt{3}$ times the nearest-neighbor carbon-carbon length,
while energies are in units of the nearest-neighbor hopping constant
$t= 2\hbar v_F/(\sqrt{3}a) \approx 3\,$eV.

\section{Wide-narrow junctions: Changing the width of a nanoribbon}

\begin{figure}
\centering
\includegraphics[clip, width = 0.7\textwidth]{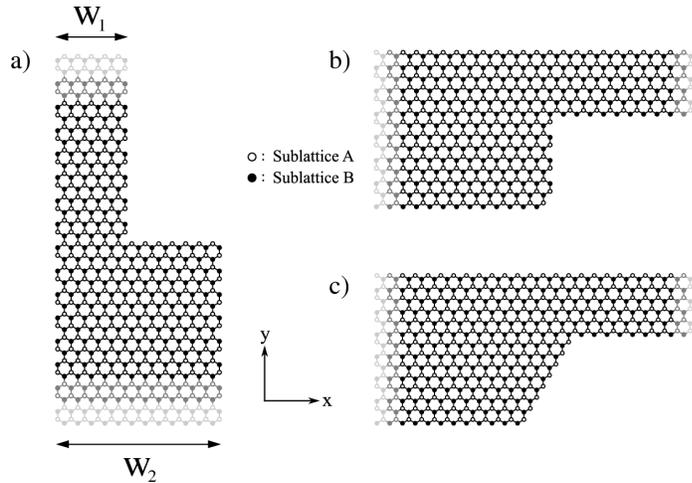}
\caption{Wide-narrow junctions for different types of
    nanoribbons formed from a hexagonal lattice. The width of the
    narrower part is $W_1$ while that of the wider part is $W_2$. The
    gray shaded sites denote infinite extension. (a)~Abrupt junction
    between armchair nanoribbons. (b)~Abrupt junction between zigzag
    nanoribbons. (c)~Gradual junction between zigzag nanoribbons.  }
\label{fig:figure1}
\end{figure}

The simplest way to form an interface
within a nanoribbon 
is to change its width. In this section we investigate the conductance of infinite
nanoribbons in which the width changes from wide to narrow, which then
can be viewed as a junction between a wide semi-infinite nanoribbon
and a narrow one. Figure \ref{fig:figure1} shows three examples of
such junctions with armchair (ac) and zigzag (zz) type edges. We
denote the width of the narrower wire by $W_1$ and the width of the
wider wire by $W_2$. A naive expectation for the dependence of $G$ on
the Fermi energy $E_F$ is the step function $G(E_F) = N_1(E_F) 2e^2/h$
where $N_1$ is the number of occupied transverse channels in the
narrow wire. This would be correct if there were no reflection at the
wide-narrow interface.
Realistically, however, there is scattering from
this interface, 
and so the steps in the conductance are not perfectly sharp.

For usual 2DEGs modeled by either a square lattice of tight-binding
sites or a continuum Schr\"odinger equation with quadratic dispersion,
the detailed shape of $G(E_F)$ has been studied previously. 
Szafer and Stone \cite{Szafer1989} calculated $G(E_F)$ by
matching the transverse modes of the two semi-infinite wires. The
inset in figure \ref{fig:figure2}\,(b) compares tight-binding results
(using a square grid) with mode-matching results in this case for $W_2
= 2W_1$ in the one-mode regime of the narrow part. The agreement
between the two is excellent. Note that the resulting conductance step
is 
very steep.

\subsection{Armchair nanoribbons}
\label{ACwidenarrow}

\begin{figure}
\centering
\includegraphics[clip, width = 0.8\textwidth]{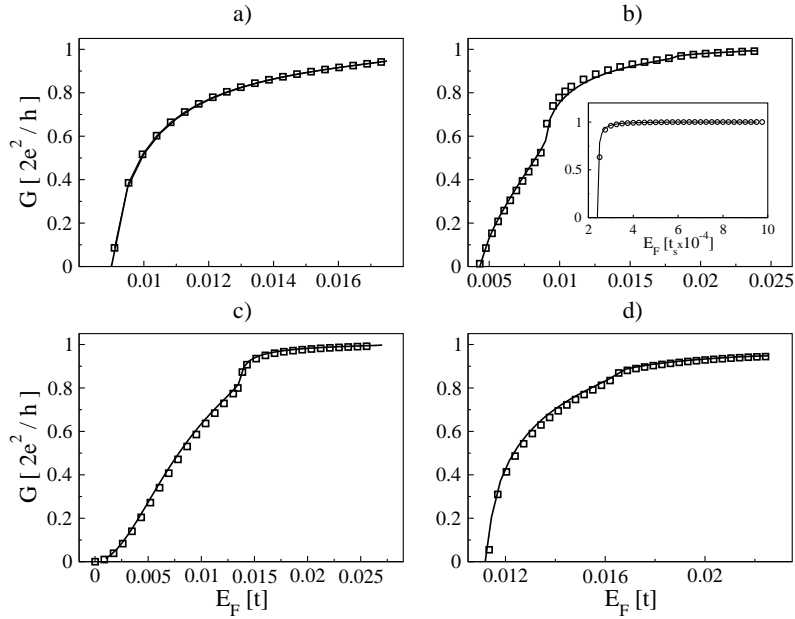}
\caption{Conductance of wide-narrow junctions in armchair
    nanoribbons as a function of Fermi energy $E_F$: results from tight
    binding (solid lines) and mode matching (squares, obtained by
    solving equation (\ref{eq:MatEqAC}) numerically). The energy window
    corresponds to the full one-mode regime of the narrow part. The
    behavior depends on whether the widths $W_{1,2}$ correspond to
    semiconducting or metallic armchair nanoribbons.  (a)
    Semiconducting-semiconducting ($W_1\!=\! 99$, $W_2\!=\!199$).  (b)
    Metallic-semiconducting ($98,199$).  (c) Metallic-metallic
    ($98,197$).  (d) Semiconducting-semiconducting ($79,109$).  Inset
    in (b): Conductance of a wide-narrow junction in a usual 2D
    electron gas: tight binding calculation (solid line) on a square
    lattice ($W_1\!=\! 200\,\mathrm{a_s}$,
    $W_2\!=\!400\,\mathrm{a_s}$) and solution of matching procedure
    (circles, equation (2) of Ref.\,\cite{Szafer1989}). $t_s$ is the
    nearest neighbor hopping energy on the square lattice and
    $\mathrm{a_s}$ is its lattice spacing.  }
\label{fig:figure2}
\end{figure}

For armchair nanoribbons, the analysis proceeds in much the same way
as for the usual 2DEG, square lattice case. At a fixed Fermi energy in
the effective Dirac equation, the transverse wavefunctions for the
various subbands are mutually orthogonal, as explained further in
\ref{app:modes}. Performing a matching procedure similar to that used
in Ref.\,\cite{Szafer1989}, one calculates the conductance from the
overlap of transverse wavefunctions on the two sides of the
wide-narrow junction. A detailed derivation is presented in
\ref{app:matching}.

Figure \ref{fig:figure2} shows the conductance resulting from the
numerical solution of the matching equations at energies for which
there is one propagating mode in the narrow part. In addition, the
conductance obtained from tight-binding calculations for wide-narrow
junctions between armchair nanoribbons is shown (using the hexagonal
graphene lattice). Figure \ref{fig:figure2} shows $G(E_F)$ for different
combinations of metallic and semiconducting nanoribbons (cf.\
\ref{app:modes}). The agreement between the two methods is extremely
good: even the singularity associated with the subband threshold in
the wider ribbon is reproduced in detail by the mode matching method,
showing that the effective Dirac equation describes the system very
well.

In figure \ref{fig:figure2}, we see immediately that $G(E_F)$ for the
armchair nanoribbon case differs greatly from the 
normal 2DEG
$G(E_F)$ [inset of figure \ref{fig:figure2}\,(b)]: the rise from zero
to unit conductance is \textit{much} slower in graphene, taking at least half
of the energy window and in some cases [see e.g.\ figure \ref{fig:figure2}\,(a)] 
not reaching the saturation value at all. For completely
metallic nanoribbons, the lineshape is very different [panel (c)] and the
conductance is suppressed at low Fermi energies (see also
reference \cite{Li2009}).

\subsection{Zigzag nanoribbons}
\label{ZZwidenarrow}

\begin{figure}
 \centering
\hspace*{-0.3cm}
\includegraphics[clip, width =0.52\textwidth]{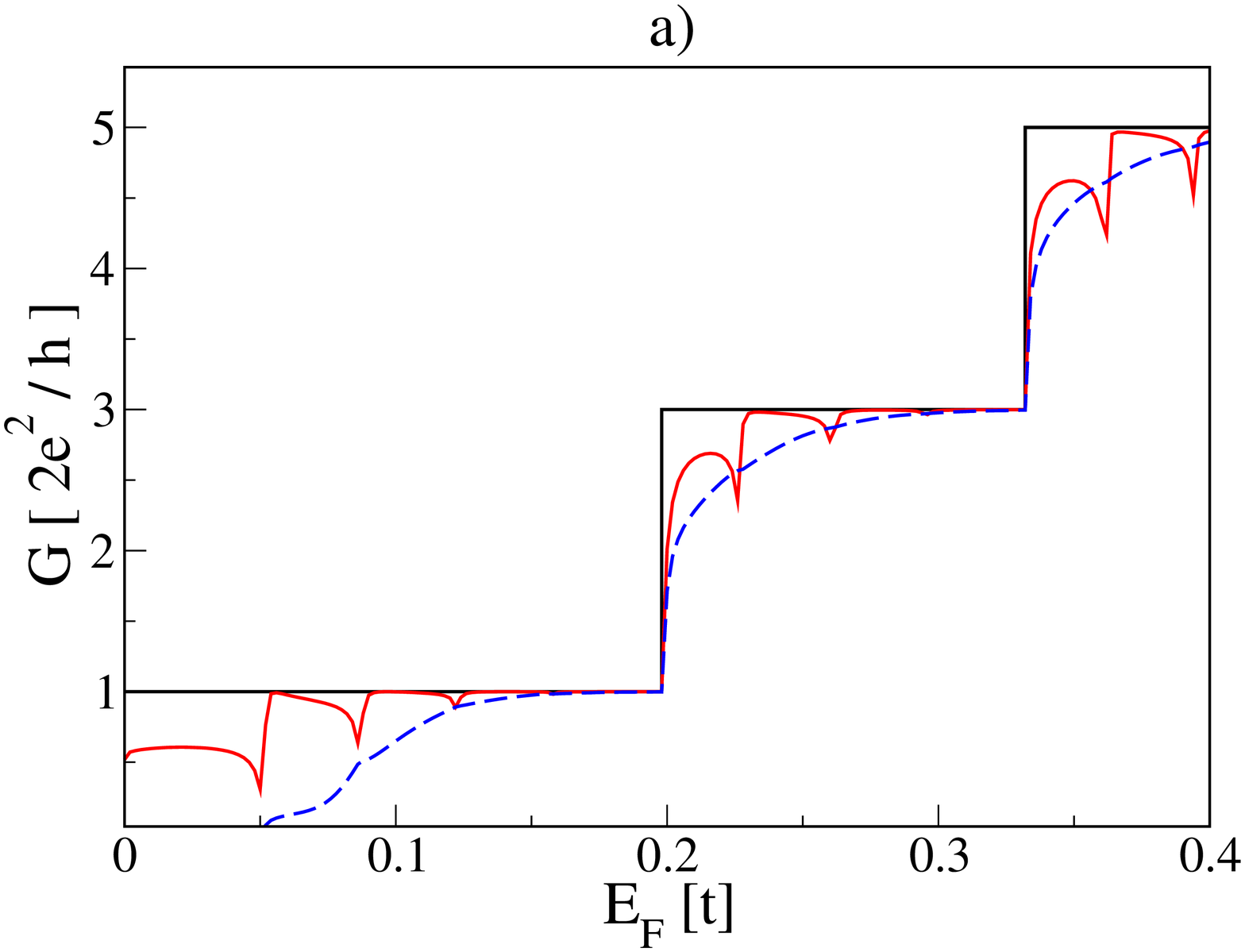}
\hspace*{0.2cm}
\raisebox{0.1cm}{\includegraphics[clip, width = 0.42\textwidth]{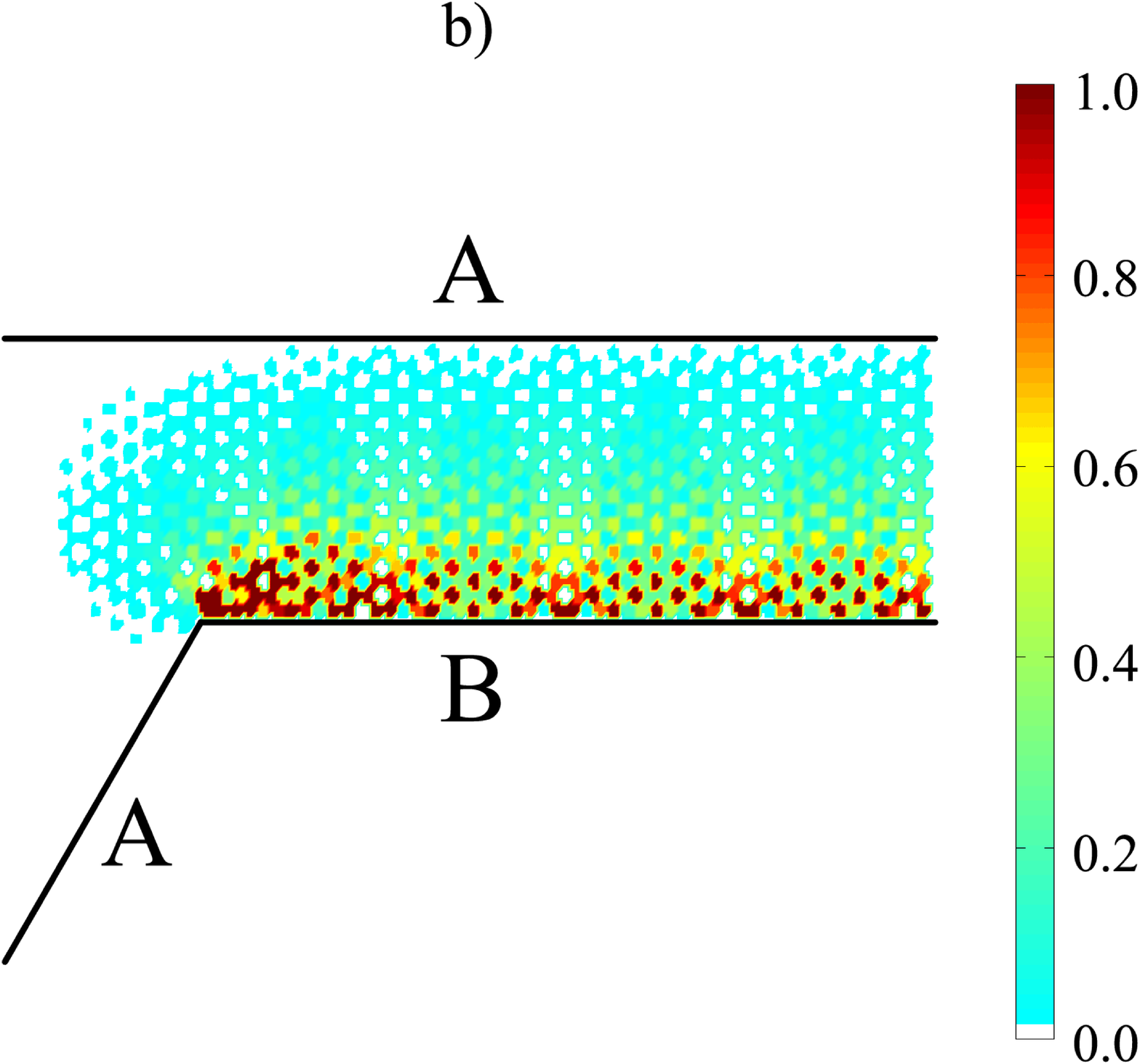}}
\caption{(a) Conductance of wide-narrow junctions in zigzag
    nanoribbons as a function of Fermi energy ($W_1 \!\approx\! 19$,
    $W_2\!\approx\!76$). Curves for a structure with abrupt change in
    width [red line, depicted in figure \ref{fig:figure1}\,(b)] and
    one with a gradually changing width [blue dashed line, depicted in
    figure \ref{fig:figure1}\,(c)] are compared to the number of
    propagating modes in the narrow wire (black line, i.e.\ the
    maximum possible conductance).  (b) Probability density
    (color-coded in arbitrary units) of an electron entering the
    system from the narrow region at $E_F\!=\!0.03\,t$. Only the
    density on the B sublattice is shown.  A and B denote the
    sublattice type at the edges. The density decreases by a factor of
    about 20 from the B edge (red) to the A edge (blue).  }
\label{fig:figure3}
\end{figure}

For zigzag nanoribbons, the analysis does not proceed as simply as in
the usual 2DEG or armchair nanoribbon cases: the transverse
wavefunctions depend on the longitudinal momentum -- similar to 2DEG wires
with a magnetic field -- and are not orthogonal at fixed Fermi energy
(cf.\ \ref{app:modes}). Because this orthogonality is used in the
matching method of \ref{app:matching}, we cannot apply it to the
zigzag case.

Figure \ref{fig:figure3}\,(a) shows numerical tight binding results
for $G(E_F)$ in two different systems with zigzag edges: one with an
abrupt change in width (red curve) and one with a gradual connection
(blue curve), as depicted in figures \ref{fig:figure1}\,(b) and (c),
respectively. Note first that the conductance is close to its maximum
value only in small windows of energy, as in the armchair nanoribbon
case and in marked contrast to the usual 2DEG, square lattice
situation.

In the abrupt case, one sees pronounced antiresonances at the
threshold energies for transverse modes in the wide nanoribbon. 
In order to see that this is due to the boundary
conditions satisfied by the transverse modes in a zigzag
nanoribbon, consider the following argument. As seen in
Figures \ref{fig:figure1}\,(b) and (c), there is
only one sublattice at each zigzag edge. In the effective Dirac
equation one has a spinor with entries corresponding to the
sublattices, thus the boundary condition is that one of the entries has to
vanish at the edge while the other component is determined by the
Dirac equation and is in general not zero at the boundary \cite{Brey2006a}. One finds
from equation (\ref{eq:quantZZ}) (e.\,g. from a graphical solution) that
the higher $E_F$ is above the threshold of a mode, the closer the 
transverse wavenumber gets to a
multiple of $\pi/W$ and the closer the value of the spinor entry in
question goes to zero. For our situation, then, the matching of a
transverse mode in the narrow nanoribbon (which is already
far above the threshold of the mode) with one in the wider
nanoribbon is particularly bad at the threshold of the latter and gets
better with increasing Fermi energy. This explains the observed
antiresonances in $G(E_F)$.

For the gradually widened junction, we insert another zigzag edge to
interpolate between the wide and narrow nanoribbon [see figure
\ref{fig:figure1}\,(c)]. In this case, the modes of the two infinitely
extended parts are not directly matched and thus the sharp
antiresonances are not present. Note, however, the complete
suppression of $G$ at very low energies. 
In this regime there is only
a single mode propagating in the wide nanoribbon as well as in the narrow one. This state is located
mainly on the B sublattice close to the lower edge and on the A sublattice on the upper edge.
Since the sublattice at the lower edge changes from A to B at the junction 
[cf. figure \ref{fig:figure1}\,(c)], this state cannot be
transmitted and the conductance is zero. This is confirmed by the
intensity distribution plotted in figure \ref{fig:figure3}\,(c).
In the more realistic next-nearest-neighbor hopping model, the situation 
is the same for most of the  single mode regime but changes for very low energies, 
when the so-called edge states are propagating \cite{Fujita1996, Brey2006a}. 
In that regime, the two states are exponentially localized at the upper and lower edge, 
respectively, and are independent of each other. Thus, the one localized at the A edge 
transmits whereas the one localized at the B edge is blocked \cite{Wimmer2008a}.

Summarizing the results for the wide-narrow junctions, we see that the
behavior of graphene nanoribbons differs substantially from that of
the familiar 2DEG situation. The matching at the graphene junctions is
much less good, leading to a suppression of the conductance from the
expected nearly step-like structure.

\section{Curved graphene nanoribbons}

\begin{figure}
\centering
\includegraphics[clip, width = 0.8\textwidth]{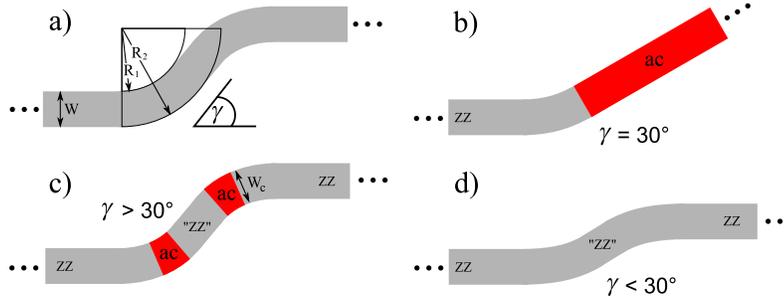}
\caption{Schematics of curved graphene nanoribbons. zz denotes
    a nanoribbon with a zigzag edge as in figure
    \ref{fig:figure1}\,(b), ac denotes an armchair edge, and ``zz''
    denotes a zigzag-like boundary as explained in the text. 
    (a)~Parameters defining a
    ``sidestep nanoribbon''; it is point symmetric about its center.
    (b)~A structure with a single zigzag-armchair interface.
    (c)~For $\gamma \!>\! 30\degr$, there will be small regions with
    armchair edges (shaded red); these have a width $\Wac$ and behave
    as in \ref{app:modes}. (d)~For $\gamma \!<\! 30\degr$, no
    armchair regions form; the curved nanoribbon is zigzag-like
    throughout.  }
\label{fig:figure4}
\end{figure}

Curved nanoribbons are defined by cutting smooth shapes out of an
infinite graphene sheet. Since the graphene lattice is discrete, the
resulting boundary is not perfectly smooth but will have edges of
zigzag and armchair type in certain directions as well as some
intermediate edge types. However, according to Akhmerov and Beenakker
\cite{Akhmerov2008a}, the intermediate boundary types behave basically
like zigzag boundaries for low energies, and we thus call these
boundaries ``zigzag-like''.

\begin{figure}
\centering
\includegraphics[clip, width = 0.7\textwidth]{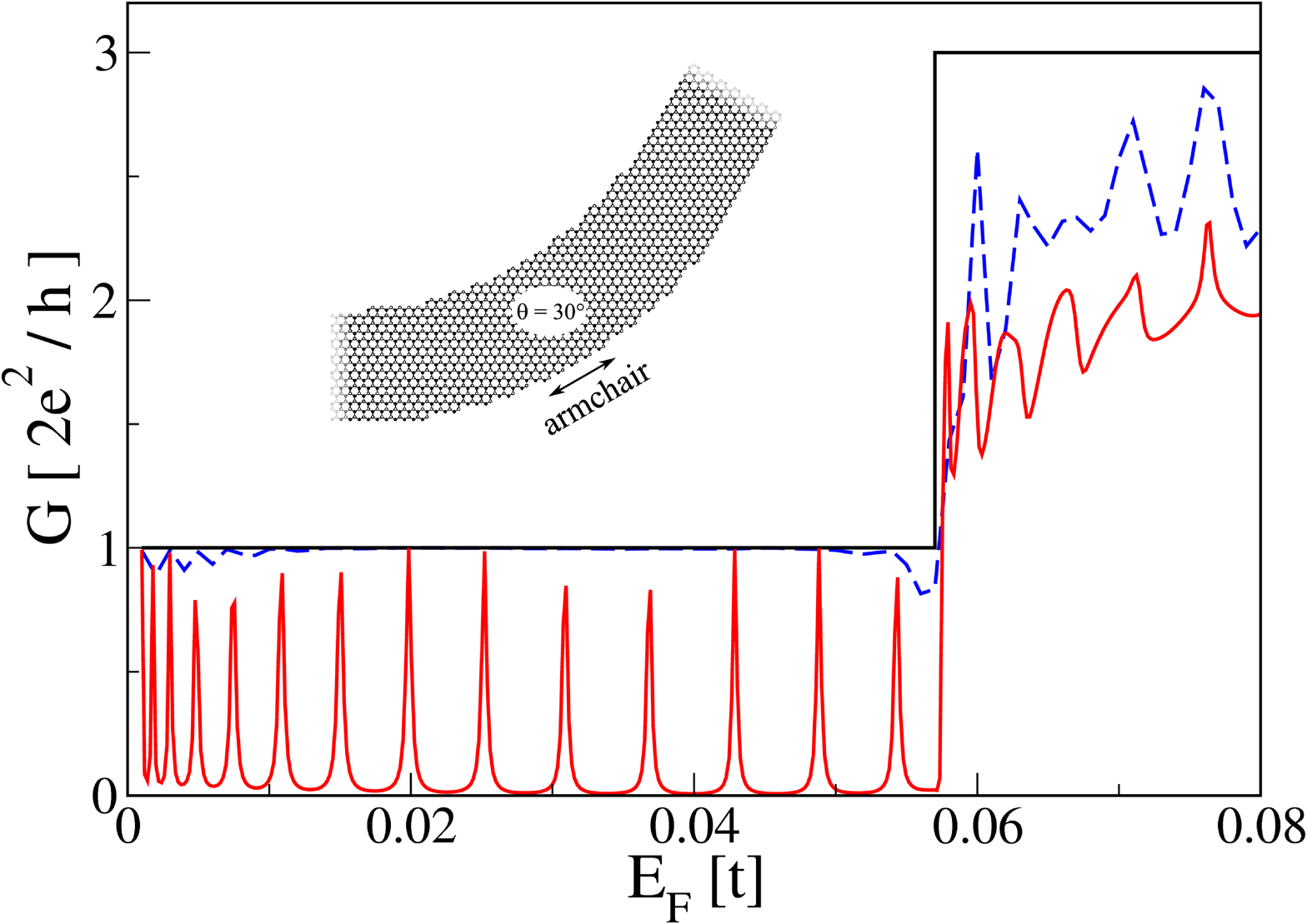}
\caption{Conductance of sidestep nanoribbons as a function of
    Fermi energy with $\gamma\!=\!60\degr$ and $R_2\!=\!259$ for two
    widths of the armchair region, $\Wac\!=\!68.5$ (solid red,
    corresponds to a semiconducting ribbon) and $\Wac\!=\!68$ (dashed
    blue, corresponds to a metallic ribbon). The solid black line
    shows the number of propagating transverse modes in the zigzag
    leads. Note that the internal interfaces between the zigzag and
    semiconducting armchair regions are much more reflective than for
    the metallic armchair case. Inset: The lattice structure of the
    first curve of a sidestep nanoribbon showing the armchair region
    formed at $\theta \!=\! 30\degr$.  }
\label{fig:figure5}
\end{figure}

In figure \ref{fig:figure4} we show schematically several of the
curved nanoribbons studied. A ``sidestep nanoribbon'' consists of an
infinitely extended horizontal zigzag ribbon of width $W$, followed by
a curved piece with outer radius of curvature $R_2$ and inner radius
$R_1 \!=\! R_2 \!-\! W$, a second straight piece making an angle
$\gamma$ with respect to the first one, a curve in the opposite
direction, and finally followed by another infinitely extended zigzag
nanoribbon. The details of the system's edge depend on $\gamma$: (1)
If $\gamma \!=\! 30\degr$, the middle straight piece has armchair
edges. (2) If $\gamma \!>\! 30\degr$ the middle straight piece is
zigzag-like with the dominating sublattice at the edges reversed from
that for the two horizontal nanoribbons. In the curved part, there is
a small region where the edges are locally of armchair type. If we
denote the angle of the local longitudinal direction from the
horizontal by $\theta$, this happens at $\theta \!=\! 30\degr$ [see figure \ref{fig:figure4}\,(c)]. 
The inset in figure \ref{fig:figure5} shows the lattice structure of such
a curved region. (3) Finally, if $\gamma \!<\! 30\degr$, the middle
straight piece also has zigzag-like edges, but the dominating
sublattice at the edges is the same as for the horizontal ribbons. In
this case, no local armchair region forms as $\theta$ is always
smaller than $30\degr$ [see figure \ref{fig:figure4}\,(d)].

In these various cases, then, different \textit{interior interfaces}
are formed between zigzag and armchair nanoribbons. We will see that
the type of interface
is critical in determining the properties of the
curved nanoribbons. In addition, the nature of the armchair nanoribbon
-- whether it is semiconducting or metallic -- has a large effect on
the conductance. Thus the width of the armchair region $\Wac$ is an
important parameter; according to equation (\ref{eq:metal}) one has a
metallic nanoribbon if $4(1\!+\!\Wac/a)/3 \in \mathbb{N}$ and a
semiconducting nanoribbon otherwise.

Figure \ref{fig:figure5} shows the conductance of sidestep nanoribbons
with $\gamma\!=\!60\degr$, for which a small armchair region is formed
in each of the curved parts. When the width of this armchair region
corresponds to a metallic nanoribbon, the conductance is basically
$2e^2/h$ -- the maximum possible value -- throughout the one-mode
regime of the zigzag leads ($\Wac\!=\!68$, dashed blue line). In
striking contrast, when the width is just $a/2$ larger (red line) the
conductance is strongly suppressed. Resonance peaks result from
Fabry-Perot behavior caused by scattering from the two armchair
regions which define a ``box'' for the middle straight region. We find
this behavior consistently for all sidestep wires in which armchair
regions form that have a width corresponding to a semiconducting
nanoribbon.

\begin{figure}
\centering
\includegraphics[clip, width = 0.6\textwidth]{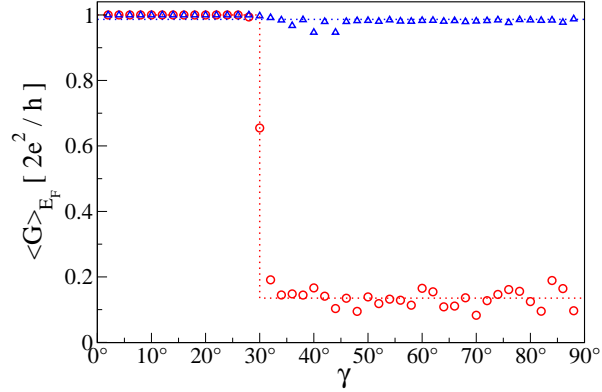}
\caption{Average conductance for two sidestep nanoribbons as a
    function of angle $\gamma$ ($R_2\!=\!259$). The average is taken
    over all Fermi energies in the one-mode regime of the zigzag
    leads. In one structure, the armchair region is metallic (blue
    triangles, $\Wac\!=\!68$) while in the other it is semiconducting
    (red circles, $\Wac\!=\!68.5$). Note the sharp decrease in
    conductance in the semiconducting case when the armchair edges
    first form at $\gamma\!=\!30\degr$. The dotted lines are guides to
    the eye.  }
\label{fig:figure6}
\end{figure}

Figure \ref{fig:figure6} shows the dependence on the angle $\gamma$ by
plotting the conductance $\langle G\rangle$ averaged over all energies for which there is
one propagating mode in the zigzag leads. For
$\gamma\!<\!30\degr$ there are no armchair regions in the curved parts
of the structure, and the average conductance is very close to the
maximum value in all cases studied. As soon as the critical angle of
$30\degr$ is surpassed and small armchair pieces form in the curves,
the conductance depends strongly on the exact value of $\Wac$. If
$\Wac$ corresponds to a metallic ribbon, $\langle G\rangle$ remains
high and is rather independent of $\gamma$. On the other hand if
$\Wac$ corresponds to a semiconducting ribbon, $\langle G\rangle$
suddenly drops by more than 80 percent and then remains approximately
constant upon further increase of $\gamma$. The constancy of $\langle
G\rangle$ in the respective regimes supports the statement of
Ref.\,\cite{Akhmerov2008a} that straight boundaries that are neither
exactly of armchair nor exactly of zigzag type behave like zigzag
boundaries. \textit{To summarize,
if a curve in a zigzag nanoribbon causes two semiconducting armchair
regions to appear, then a very effective barrier is formed which
causes very high reflectivity.}

\begin{figure}
\centering
\includegraphics[clip, width = 0.82\textwidth]{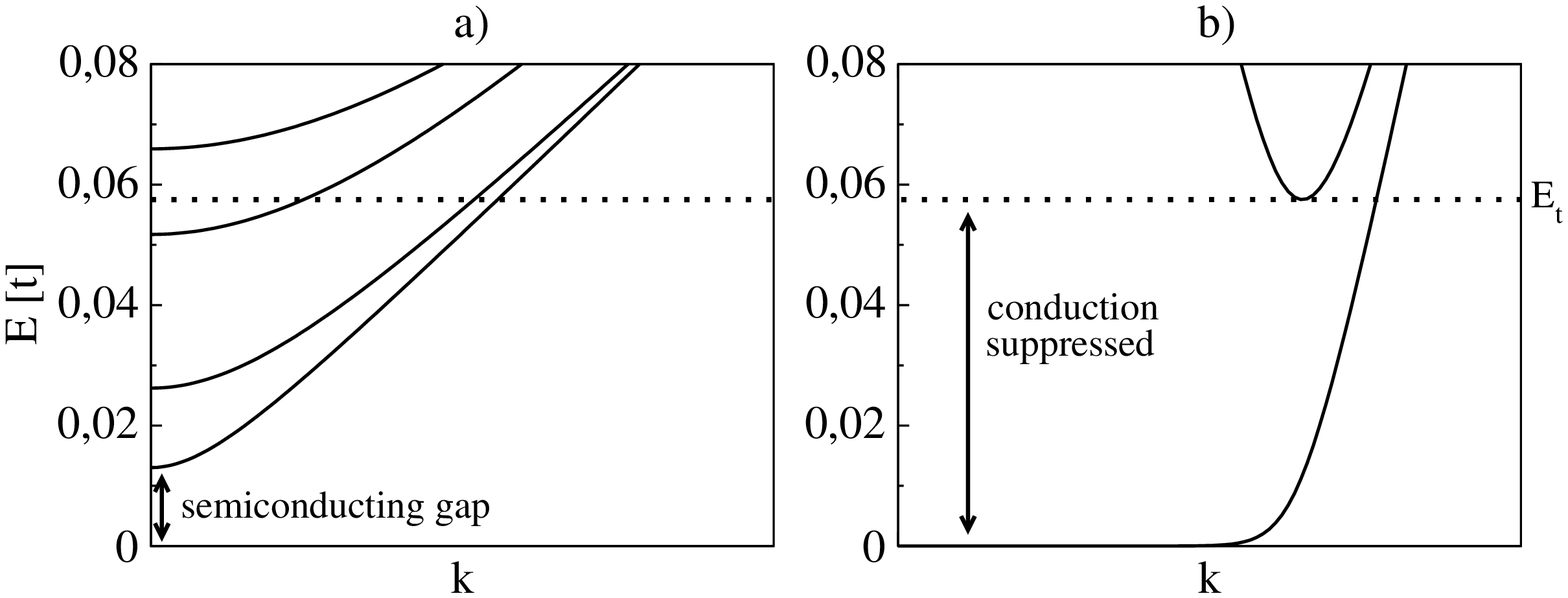}
\caption{ Tight binding band structures of infinitely extended
    graphene nanoribbons. (a)~Armchair nanoribbon with a width of
    $68.5$ (same width as the local armchair piece forming in the structure of figure \ref{fig:figure5}, red curve). 
    (b)~Zigzag nanoribbon with a width of
    $(39\!+\!\frac{1}{2})\sqrt{3} \approx 68.4$. The semiconducting
  energy gap in the armchair nanoribbon does not correspond to the
  energy region in figure \ref{fig:figure5} in which the conductance is suppressed.
  }
\label{fig:figure7}
\end{figure}

The simplest idea to explain this effect would be that at low energies
there is by definition a gap in a semiconducting ribbon and since this
means there are no propagating states in the local armchair region,
one expects the conductance to be suppressed because electrons have to
tunnel trough this region in order to be transmitted. However, this
does \textit{not} explain our findings: the energy range over which
the conductance suppression occurs is much larger than the
energy gap of the semiconducting region. In fact it is given by the energy range of
the one-mode regime in the surrounding zigzag parts. To make this
clear, we show the bandstructures of both a semiconducting armchair
ribbon and a zigzag ribbon of approximately the same width in figure
\ref{fig:figure7} (both nanoribbons are infinitely extended). One can
clearly see that within the one-mode regime of the zigzag nanoribbon,
in which the states are completely valley polarized,
there can be several propagating modes in the semiconducting armchair
nanoribbon, so the suppression of $G$ must be of a different origin.

\begin{figure}
\centering
\includegraphics[clip, width = 0.5\textwidth]{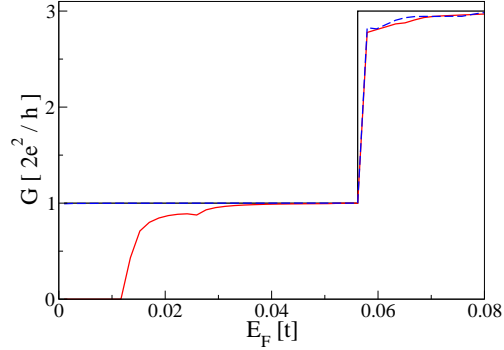}
\caption{ Single zigzag to armchair interface 
conductance of a
    smooth bend through $30\degr$, as depicted in figure
    \ref{fig:figure4}\,(b) ($R_2\!=\!259$). Both semiconducting (solid
    red line, $\Wac\!=\!68.5$) and metallic (dashed blue line,
    $\Wac\!=\!68$) armchair nanoribbons lead to good conductance. The solid black line
    shows the number of propagating transverse modes in the zigzag lead, corresponding to the maximum 
    possible conductance (in the armchair lead at energies above the semiconducting
gap there are always more or equally many modes propagating). }
\label{fig:figure8}
\end{figure}

Furthermore, it is \textit{not} the bare zigzag-armchair junction that leads to
suppressed conductance, but rather it is necessary to have two zigzag
pieces differing by an angle of more than $30\degr$ and being
separated by a small armchair region. This can be seen in two
stages. First, figure \ref{fig:figure8} shows the conductance of an
infinitely extended zigzag nanoribbon connected to an infinitely
extended armchair nanoribbon via a $30\degr$ curve, the structure
shown schematically in figure \ref{fig:figure4}\,(b). In the one-mode
regime of the zigzag ribbon, the conductance is maximal for the case
of a metallic armchair ribbon. For a semiconducting
armchair ribbon, the conductance is, of course, zero for energies
below the gap, but it increases rapidly up to $2e^2/h$ for
larger values of $E_F$. \textit{Thus, a single zigzag to
semiconducting-armchair interface
conducts well.}

\begin{figure}
\centering
\includegraphics[clip, width = 0.65\textwidth]{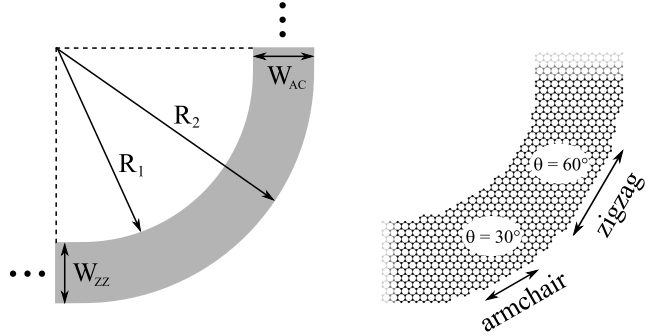}
\caption{ $90\degr$ curve with horizontal zigzag lead and
    vertical armchair lead. A local armchair region forms at
    $\theta\!=\!30\degr$ and a local zigzag region at
    $\theta\!=\!60\degr$.  }
\label{fig:figure9}
\end{figure}

\begin{figure}
\centering
\includegraphics[clip, width = 0.9\textwidth]{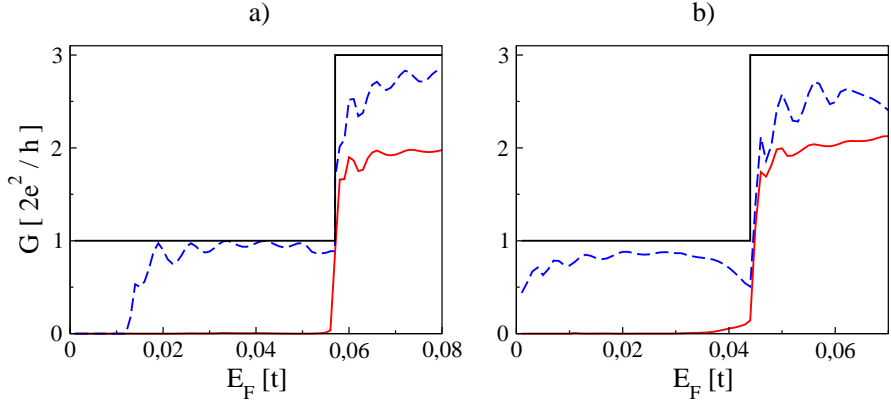}
\caption{Conductance of $90\degr$ curved nanoribbons with
    either a semiconducting (solid red) or metallic (dashed blue)
    armchair region at $\theta\!=\!30\degr$.  (a)~Semiconducting
    armchair lead: $\Wac\!=\!69$ (solid red) and $\Wac\!=\!69.5$
    (dashed blue).  (b)~Metallic armchair lead: $\Wac\!=\!88.5$ (solid
    red) and $\Wac\!=\!89$ (dashed blue).  Black line: number of
    propagating modes in the zigzag lead. ($R_2\!=\!259$.)  }
\label{fig:figure10}
\end{figure}

\begin{figure}[t]
 \centering
\includegraphics[clip, width = 0.6\textwidth]{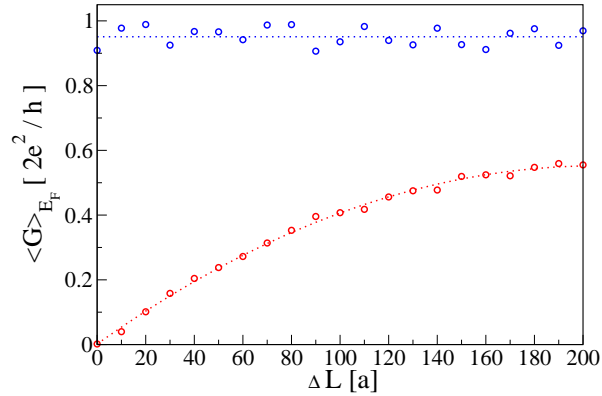}
\caption{Average conductance as a function of the length added
    to the armchair region of a $90\degr$ curved nanoribbon. The
    structures are as in figure\,\ref{fig:figure10}\,(a): nanoribbons
    with a semiconducting (red) or metallic (blue) armchair region at
    $\theta\!=\!30\degr$. The average is over all $E_F$
    above the semiconducting gap of the armchair lead and in the one
    mode regime of the zigzag lead. The dotted lines are guides to the
    eye.}
\label{fig:figure11}
\end{figure}

For the second stage of the argument, consider a bend through
$90\degr$ from an infinite zigzag lead to an armchair one, as depicted
in figure \ref{fig:figure9}. In contrast to the $30\degr$
zigzag-armchair connection just discussed, this one has three interfaces
between zigzag and armchair regions. Figure
\ref{fig:figure10} shows the conductance of several $90\degr$ curved
nanoribbons. As for the sidestep ribbons, the conductance is
suppressed when a semiconducting armchair region is present in the
curve. Note that the suppression is \textit{not} due to the infinitely
extended armchair lead, for which we chose a semiconducting nanoribbon
in \ref{fig:figure10}\,(a) and a metallic one in
\ref{fig:figure10}\,(b).

If one makes the natural assumption that the armchair region acts as a
blocking barrier, one would expect the blocking to become more
effective as the armchair region is lengthened. However, this is
clearly \textit{not} the case in the data shown in figure
\ref{fig:figure11}. The system is a $90\degr$ curved nanoribbon in
which the armchair region at $\theta=30\degr$ is lengthened by $\Delta
L$; we plot $\langle G\rangle$, the conductance averaged over all
energies in the one-mode regime of the zigzag lead, as a function of
$\Delta L$. For a metallic armchair region in the curve, the
conductance is roughly independent of $\Delta L$, as
expected. Surprisingly, for the semiconducting case, the conductance
\textit{increases} as a function of $\Delta L$.
This establishes,
then, that \textit{conductance suppression occurs when two zigzag-armchair
interfaces occur in close spatial proximity.}

Our numerical results 
suggest that the evanescent modes in the armchair
regions play an essential role. They are necessary, of course, in
order to match the propagating zigzag mode to a solution in the
armchair region. For short armchair pieces the evanescent modes from
the two interfaces 
overlap. We conjecture that these evanescent modes
are mutually incompatible in the semiconducting case, destroying the
possibility of matching on both sides at the same time, while they are
compatible for metallic armchair regions. If one has a long armchair
piece, the evanescent modes decay leading to independent matching at
the two ends.

\section{Conclusions}

\begin{figure}
\centering
\includegraphics[clip, width = 0.7\textwidth]{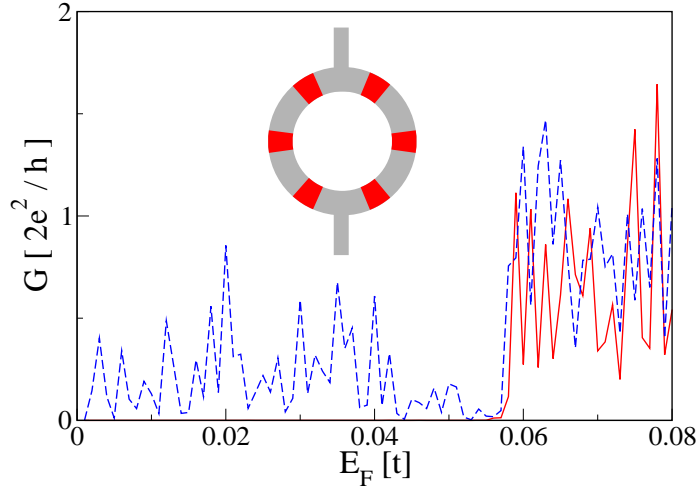}
\caption{Conductance of rings with armchair leads
    ($R_2\!=\!259$).  Red solid: Ring with semiconducting armchair
    regions in both arms ($W_c\!=\!69$ in the right arm and
    $W_c\!=\!68.5$ in the left).  Blue dashed: Ring with metallic
    armchair regions in the right arm ($W_c\!=\!69.5$) and
    semiconducting in the left ($W_c\!=\!69$).  Inset: Schematic of
    the ring structure; red shading indicates regions with armchair
    edges, as in figure \ref{fig:figure4}.  }
\label{fig:figure12}
\end{figure}

We have shown in a variety of examples that interfaces
within graphene nanoribbons can strongly affect their conductance, much more so than
in the familiar 2DEG electron waveguides and wires. First, for
wide-narrow junctions, our main results are figures \ref{fig:figure2}
and \ref{fig:figure3}. For both armchair and zigzag nanoribbons,
changes in width act as a substantial source of scattering, reducing
the conductance. Second, for curved nanoribbons, our main results are
figures \ref{fig:figure6}, \ref{fig:figure8}, and
\ref{fig:figure11}. There is a strong reduction in conductance when a
curve joining two zigzag regions contains a semiconducting armchair
region.

The effect of such internal interfaces will certainly be felt in more
complex structures as well. As an example, consider rings for studying
the modulation of the conductance by a magnetic field through the
Aharonov-Bohm effect \cite{Wurm2009a}. Figure \ref{fig:figure12} shows
such a ring schematically together with its conductance in two
cases. As for the curved nanoribbons, when semiconducting armchair regions
occur in the curved part of the structure, the conductance is
substantially reduced.

In considering experimental manifestations of internal interfaces, disorder, 
and in particular the edge disorder which has received attention 
recently \cite{Jia2009, Girit2009, Areshkin2007, Evaldsson2008, Cresti2009, Martin2009}, 
may be important. The effects we observe in our calculations will most likely 
also be present in structures with disordered edges, provided the disorder is 
not too strong. Consider, for example, the suppression of the conductance in 
curved wires. In the inset of figure \ref{fig:figure5} as well as in figure 
\ref{fig:figure9}, one can see that between the armchair and the zigzag regions, 
the edges are not perfect. We believe that when the edge disorder is weak enough 
to allow for pieces with armchair edges, the suppression should still be present.

The underlying reason for the impact of internal interfaces
can be viewed in two ways. From the lattice point of view, it arises from 
the additional complexity of the hexagonal lattice with its basis compared
to the standard square lattice. Equivalently, from the continuum point
of view, it arises from the extra degrees of freedom inherent in the
Dirac-like equation governing graphene -- those of the sublattice and
valley pseudospins. As development of graphene nanostructures
accelerates, the impact of internal interfaces
should be taken into account when considering future carbon nanoelectronic schemes.

\ack 
We would like to thank Adam Rycerz for valuable discussions.
JW acknowledges support from Deutsche Forschungsgemeinschaft
within GRK 638, and MW, IA and KR support from Deutsche
Forschungsgemeinschaft within SFB 689. IA is supported by the funds 
of the Erdal \.In\"on\"u Chair of Sabanc\i~University. The work at Duke was supported
in part by the U.S.\,NSF (DMR-0506953) and the DAAD.

\appendix

\section{Wavefunctions of graphene nanoribbons in the Dirac equation}
\label{app:modes}

\begin{figure}[b]
\centering
\includegraphics[width = 0.8\textwidth]{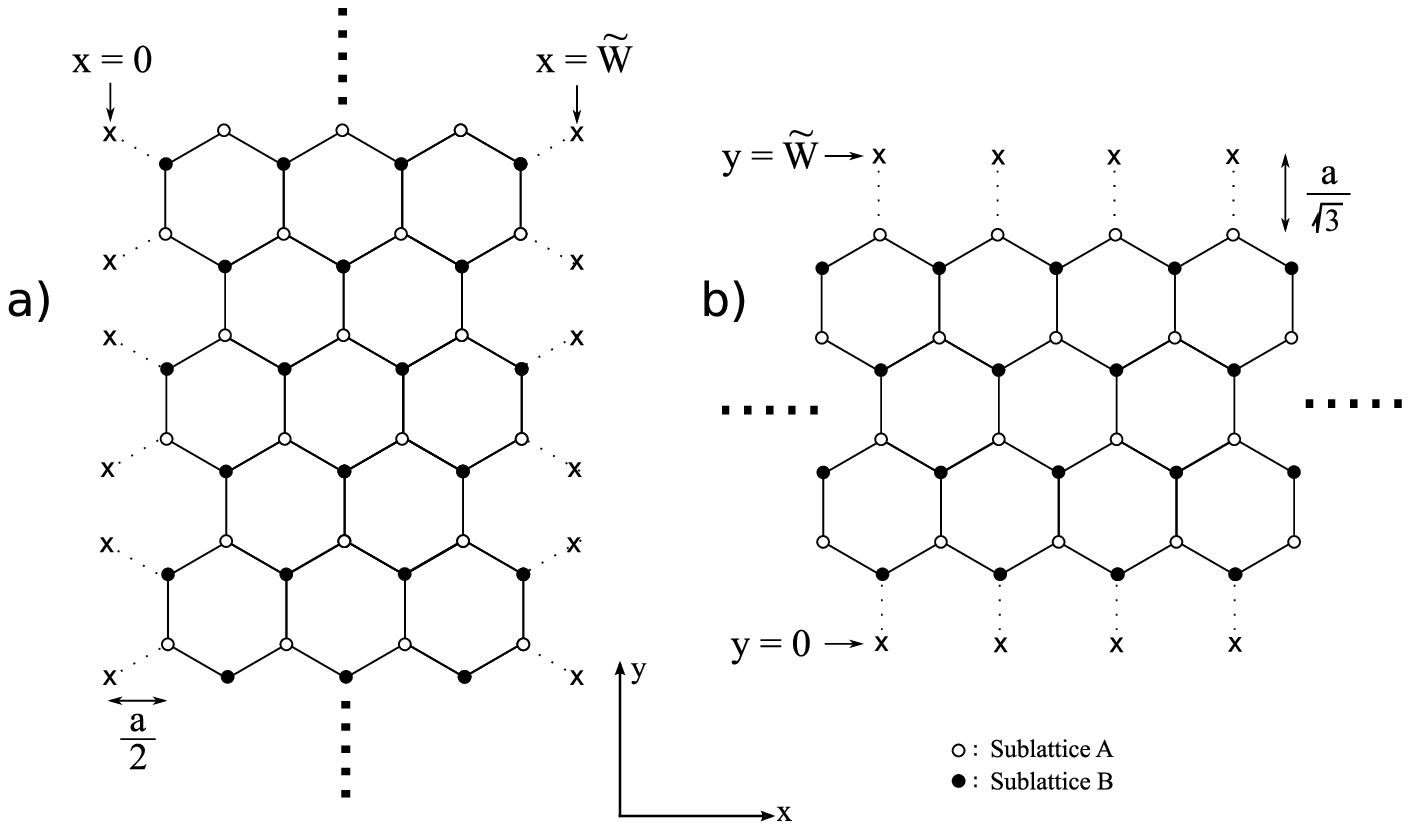}
\caption{Infinitely extended graphene nanoribbons. (a)
    Armchair ribbon along the $y$-direction. The outermost rows of
    atoms are at $x=a/2$ and $\tilde{W}-a/2$ respectively. Hence, the
    width of the ribbon is given by $W=\tilde{W}-a$. The boundary
    condition however is, that the wavefunction is zero at $x=0$ and
    $x=\tilde{W}$ respectively.  (b) Zigzag ribbon along the
    $x$-direction. Here the width of the ribbon is
    $W=\tilde{W}-2a/\sqrt{3}$. Since first row of missing atoms at
    each side is only on one sublattice, the boundary conditions
    requires only the corresponding part of the wavefunction to
    vanish.  }
\label{fig:figureA1}
\end{figure}

We calculate the eigenfunctions of graphene nanoribbons withing the
effective Dirac model. This has been done by Brey and Fertig in
\cite{Brey2006a} and Peres, \etal in \cite{Peres2006a}.  The
effective Dirac equation taking into account contributions from both
valleys is given by \cite{CastroNeto2009}
\begin{equation}
\label{eq:dirac}
 H \Phi(\bsy{r}) = E \Phi(\bsy{r})
\end{equation}
with 
\begin{equation}
H = v_F
\left(
\begin{array}{cc}
 \sigma_x p_x+\sigma_y p_y & 0 \\
0 & -\sigma_x p_x + \sigma_y p_y
\end{array}
\right)
\end{equation}
and
\begin{equation}
\Phi(\bsy{r}) = \left[\Phi_K(\bsy{r}),\Phi_{K'}(\bsy{r})\right]^T =  \left[\Phi_A(\bsy{r}), \Phi_B(\bsy{r}), \Phi_{A}'(\bsy{r}),\Phi_{B}'(\bsy{r}) \right]^T \,.
\end{equation}
Here $\Phi_{K}$ and $\Phi_{K'}$ are spinors with two components corresponding to contributions from the two different valleys $K$ and $K'$ respectively. $\Phi_{A/B}$ and $\Phi_{A/B}'$ are scalar wavefunctions, where the subscripts $A$ and $B$ stand for the two sublattices
(see figure \ref{fig:figureA1}).
The total wavefunction containing the fast oscillations from the $K$-points is then
\begin{equation}
\label{eq:fullpsi}
 \psi(\bsy{r}) = {
  \psi_A(\bsy{r})  \choose \psi_B(\bsy{r}) 
 }
=e^{i\bsy{K}\cdot\bsy{r}} 
{
 {\Phi}_A(\bsy{r}) \choose
 {\Phi}_B(\bsy{r})} +
e^{i\bsy{K}'\cdot\bsy{r}} 
{
 {\Phi}_A'(\bsy{r}) \choose
 {\Phi}_B'(\bsy{r})
}\,.
\end{equation}

\subsection{Armchair nanoribbons}
We consider an armchair nanoribbon which is infinitely extended along the $y$-direction [see figure \ref{fig:figureA1}\,(a)].
Using the Bloch ansatz
\begin{equation}
 \Phi(\bsy{r}) = e^{ik_yy}\phi(x)
\end{equation}
and the Dirac equation (\ref{eq:dirac}), one obtains
\begin{eqnarray}
\label{eq:DiracAC_1}
  -i(k_y+\partial_x)\phi_B(x) = \epsilon \phi_A(x) \\
\label{eq:DiracAC_2}
  i(k_y-\partial_x)\phi_A(x) = \epsilon \phi_B(x) \\
\label{eq:DiracAC_3}
  -i(k_y-\partial_x)\phi_B'(x) = \epsilon \phi_A'(x) \\
\label{eq:DiracAC_4}
  i(k_y+\partial_x)\phi_A'(x) = \epsilon \phi_B'(x) 
\end{eqnarray}
and, by applying the Hamiltonian twice,
\begin{equation}
\label{eq:twiceH_AC}
 (k_y^2-\partial_x^2)\phi(x) = \epsilon^2\phi(x)
\end{equation}
with $\epsilon = E/(\hbar v_F)$. According to figure \ref{fig:figureA1}\,(a), the correct boundary condition \cite{Brey2006a} for an armchair nanoribbon is $\psi(\bsy{r})=0$ 
for $x=0$ and $x=\tilde{W}$. (For the connection between
the nanoribbon width $W$ used previously and $\tilde{W}$, see the caption
of figure \ref{fig:figureA1}.) 
The ansatz
$\phi_B(x) = Ae^{iq_nx}+Be^{-iq_nx}$, $\phi_B'(x) = Ce^{iq_nx}+De^{-iq_nx}$
solves both the B sublattice parts of equation (\ref{eq:twiceH_AC}) with $\epsilon^2=k_y^2+q_n^2$ and the boundary condition, if we require 
\begin{equation}
\label{eq:qn}
 q_n = \frac{n\pi}{\tilde{W}} - K \quad \mathrm{with~} n\in \mathbb{Z}
\end{equation}
where $K=4\pi/(3a)$. We find that $B=C=0$ and $A=-D$.
Using equations (\ref{eq:DiracAC_1}) and (\ref{eq:DiracAC_3}) to determine $\Phi_A(x)$ and $\Phi_A'(x)$ from $\Phi_B(x)$ and $\Phi_B'(x)$, we thus find that, up to a normalization factor, the wavefunctions are 
\begin{equation}
\label{eq:phiac}
\fl \phi(x) \sim \left[ (q_n-ik_y)e^{iq_nx}/\epsilon,\; e^{iq_nx},\; -(q_n-ik_y)e^{-iq_nx}/\epsilon,\; -e^{-iq_nx}\right]^T\,,
\end{equation}
\begin{equation}
\label{eq:psiac}
 \psi(\bsy{r}) \sim e^{ik_yy}\sin\left[(q_n+K)x\right] \left[(q_n-ik_y)/\epsilon,\; 1 \,\right]^T\,.
\end{equation}

The wavefunction $\psi(\bsy{r})$ is, up to the spinor part, very similar to that of a 2DEG waveguide:
the width of the ribbon is a multiple of half the transverse wavelength. However, here the transverse wavelength is of order the 
lattice constant, not the system's width, since $n$ is of order $\tilde{W}/a$ for the energetically lowest lying modes. Nevertheless,
the wavefunctions for different transverse quantum numbers $n$ are orthogonal at a fixed Fermi energy.
Note that for evanescent modes we just have to consider imaginary wavenumbers $k_y=i\kappa_y$ and
equations (\ref{eq:qn}), (\ref{eq:phiac}), and (\ref{eq:psiac}) still hold.

The energy for this solution is
\begin{equation}
 E=\pm\hbar v_F \sqrt{k_y^2+q_n^2}\,.
\end{equation}
Therefore one has a metallic spectrum if there is a state with $q_n=0$. From equation (\ref{eq:qn}) it follows immediately that this is
the case whenever 
\begin{equation}
 \label{eq:metal}
 \frac{4}{3}\frac{\tilde{W}}{a} \in \mathbb{N}\,.
\end{equation}

\subsection{Zigzag nanoribbons}

For a zigzag nanoribbon along the $x$-direction [see figure \ref{fig:figureA1}\,(b)], the Bloch ansatz is
\begin{equation}
 \Phi(\bsy{r}) = e^{ik_xx}\phi(y) \;,
\end{equation}
the Dirac equation becomes
\begin{eqnarray}
\label{eq:DiracZZ_1}
  (k_x-\partial_y)\phi_B(y) = \epsilon \phi_A(y) \\
\label{eq:DiracZZ_2}
  (k_x+\partial_y)\phi_A(y) = \epsilon \phi_B(y) \\
\label{eq:DiracZZ_3}
  -(k_x+\partial_y)\phi_B'(y) = \epsilon \phi_A'(y) \\
\label{eq:DiracZZ_4}
  -(k_x-\partial_y)\phi_A'(y) = \epsilon \phi_B'(y) \;,
\end{eqnarray}
and one has
\begin{equation}
\label{eq:twiceH_ZZ}
 (k_x^2-\partial_y^2)\phi(y) = \epsilon^2\phi(y) \;.
\end{equation}

The boundary condition for a zigzag ribbon differs from that for an armchair ribbon in that the wavefunction has to vanish on only one sublattice at
each edge \cite{Brey2006a}: $\psi_A(x,y=0)=\psi_B(x,y=\tilde{W}) = 0$. With the following ansatz for $\Phi_A(x)$ and $\Phi_A'(x)$,
\begin{equation}
 \phi_A(y) = Ae^{izy}+Be^{-izy} \quad\quad \phi_A'(y) = Ce^{izy}+De^{-izy}\,,
\end{equation}
(\ref{eq:twiceH_ZZ}) yields $\epsilon^2=k_x^2+z^2$, and the boundary condition requires
$A=-B$ and $C=-D$.
Thus the valleys completely decouple for zigzag nanoribbons, 
and equations (\ref{eq:DiracZZ_2}) and (\ref{eq:DiracZZ_4}) yield
\begin{equation}
\label{eq:PhiZZ}
 \Phi_{K/K'} \sim \left[\sin(zy),\; \left\{\tau\,k_x\sin(zy)+z\cos(zy)\right\}/\epsilon\,\right]^T
\end{equation}
where $\tau=+1$ for the $K$ and $\tau=-1$ for the $K'$ valley. 
The boundary condition for the $B$ parts of the wavefunction provides
an equation that determines the allowed values for $z$,
\begin{equation}
\label{eq:quantZZ}
 k_x = -\tau z / \tan(z\tilde{W})\,.
\end{equation}
Thus the transverse quantum number is coupled to the longitudinal momentum, as in 2DEG waveguides in the presence of a magnetic field.
In order to write equation (\ref{eq:PhiZZ}) in a symmetric way, we square the quantization condition (\ref{eq:quantZZ}) and use the relation 
$k_x^2 = \epsilon^2 - z^2$ to obtain 
\begin{equation}
 \label{eq:quantZZ_sq}
\epsilon^2 = z^2 / \sin^2(z\tilde{W})\,.
\end{equation}
Using (\ref{eq:quantZZ}) and (\ref{eq:quantZZ_sq}) in equation (\ref{eq:PhiZZ}) leads to 
\begin{equation}
\label{eq:PhiZZ_sym}
 \Phi_{K/K'} \sim \left[\sin(zy),\; s(z,\epsilon)\sin\{z(W-y)\}\right]^T
\end{equation}
with
%
$s(z,\epsilon) = \mathrm{sign} [\epsilon z / \sin(z\tilde{W})]$.
>From this symmetric expression, one clearly sees that the total weight on each sublattice is the same.

\begin{figure}
\centering
\includegraphics[width = 0.4\textwidth]{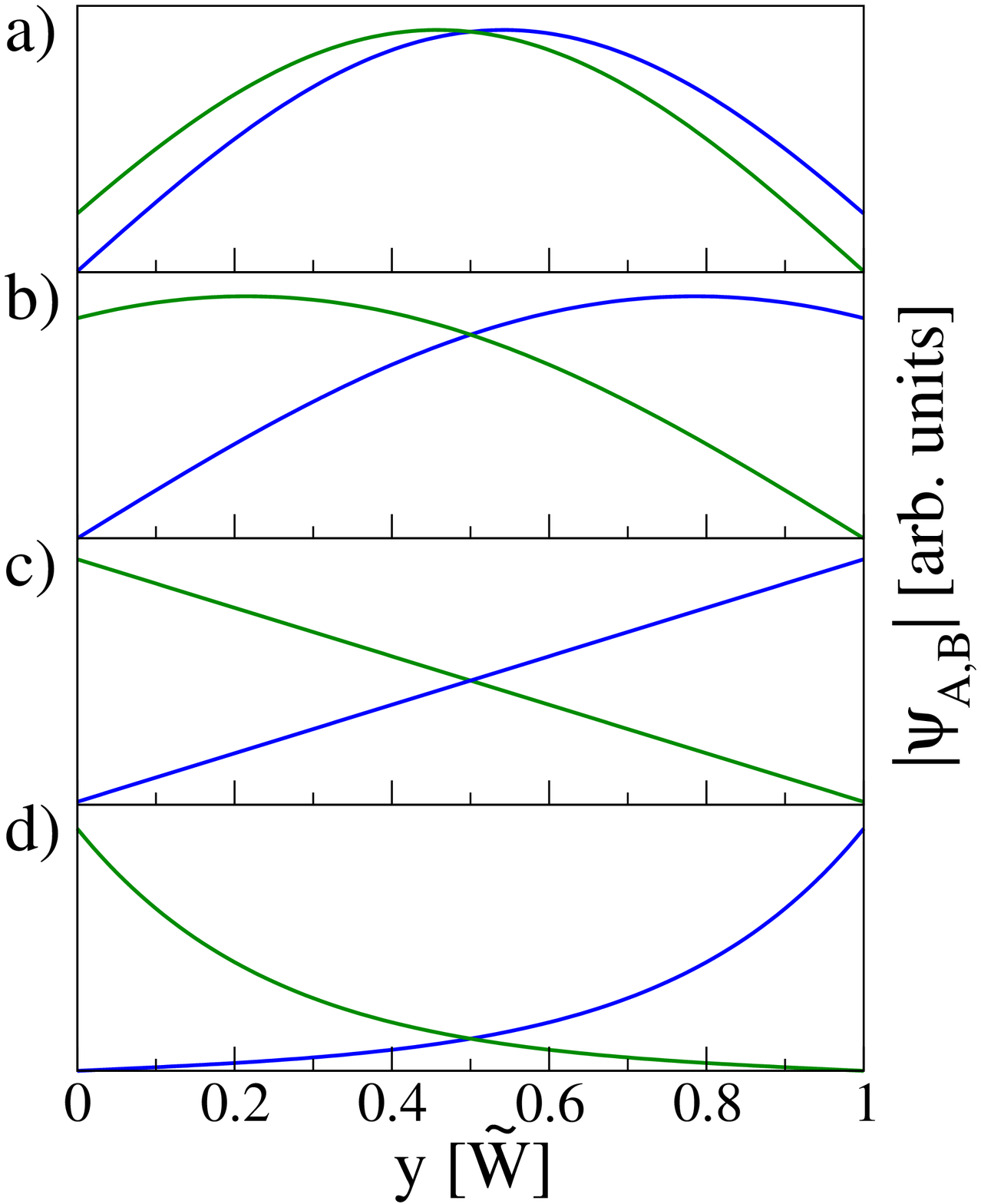}
\caption{Profile of transverse wavefunctions in a zigzag nanoribbon. (a) $z=2.25/\tilde{W}, ~\epsilon= 2.89/\tilde{W}$.
(b) $z=2/\tilde{W}, ~\epsilon= 2.2/\tilde{W}$. (c) $z=0, ~\epsilon= 1/\tilde{W}$. (d) $z=4i/\tilde{W}, ~\epsilon= 0.15/\tilde{W}$.}
\label{fig:figureA2}
\end{figure}

The transcendental equation (\ref{eq:quantZZ}) has real solutions $z \in \mathbb{R}$ only for $|\epsilon| \geq 1/\tilde{W}$. These states correspond to bulk states: they are extended over
the whole width of the ribbon. For $|\epsilon| < 1/\tilde{W}$ there are only imaginary solutions $iz\in \mathbb{R}$, corresponding to the so-called edge states \cite{Fujita1996, Brey2006a},
which are exponentially localized at the edges and live predominantly on one sublattice at each side, as can be seen from equation (\ref{eq:PhiZZ}).
For the special case $z= 0$  corresponding to $|\epsilon| = 1/\tilde{W}$ equation (\ref{eq:PhiZZ_sym}) results in 
\begin{equation}
 \lim_{|z|\rightarrow 0}\Phi_{K/K'} \sim [y,\; -\mathrm{sign}(\epsilon)(y-\tilde{W})]^T\,,
\end{equation}
i.\,e.\ a linear profile of the transverse wavefunction.
Figure (\ref{fig:figureA2}) shows the profile of several transverse zigzag modes.

\section{Mode matching for wide-narrow junctions with armchair edges}
\label{app:matching}

We derive a set of analytic equations that determine the transmission amplitudes for wide-narrow junctions with armchair edges as introduced in section \ref{ACwidenarrow}.
We label the transverse modes in the narrow part of the system by $\varphi^{\pm}(x)$ and those in the wide part by $\chi^{\pm}(x)$. The $\pm$ stands for propagation
in positive and negative $y$-direction respectively. Furthermore we use latin subscripts $n$ and $m$ for $\varphi$ and greek subscripts $\nu$ and $\omega$ for $\chi$. 
Then we know from \ref{app:modes} that
\begin{eqnarray}
 \varphi^{\pm}_n(x) = \frac{1}{\sqrt{W_1}} \sin\left(n\pi x/W_1\right)\left[(q_n\mp ik_y^n)/\epsilon,\; 1\,\right]^T \\
 \chi^{\pm}_{\nu}(x) = \frac{1}{\sqrt{W_2}} \sin\left(\nu\pi x/W_2\right)\left[(q_\nu\mp ik_y^\nu)/\epsilon,\; 1\,\right]^T\,.
\end{eqnarray}
Here, we define the $k_y^{n/\nu}$ to lie always on the positive real axes for propagating states and on the positive imaginary axes for evanescent states
\begin{equation}
k_y^{n/\nu} = +\sqrt{\epsilon^2-q^2_{n/\nu}}\,.
\end{equation}

The full scattering wavefunction for an electron incident from the wide side in mode 
$\omega$ is
\begin{eqnarray}
  y \leq 0: \quad \psi_{\omega}(x,y) &=& \chi_{\omega}^+(x)e^{ik_y^{\omega}y} + \sum_{\nu}r_{\nu\omega}\chi_{\nu}^-(x)e^{-ik_y^{\nu}y}
\\
  y \geq 0: \quad \psi_{\omega}(x,y) &=& \sum_{n}t_{n\omega}\varphi_{n}^+(x)e^{ik_y^{n}y}\,,
\end{eqnarray}
where the sums run over all modes, both propagating and evanescent. Matching the two parts at the junction, defined to be $y=0$, we obtain
\begin{equation}
\label{eq:match}
\chi_{\omega}^+(x) + \sum_{\nu}r_{\nu\omega}\chi^-_{\nu}(x) = \sum_nt_{n\omega} \varphi^+_n(x)\,.
\end{equation}
We can extract the scattering amplitudes by projecting this equation first on the wide side and then on the narrow side. First,
multiplying the B-part of this equation by $\left[\chi_{\nu',B}^-(x)\right]^*$ and integrating from $0$ to $W_2$ yields
\begin{equation}
\label{eq:reflect}
 r_{\nu\omega} = -\delta_{\nu\omega} + \sum_n2t_{n\omega} b_{\nu n}^{-+}
\end{equation}
for which we used $\int_0^{W_2}dx\left[\chi_{\nu',B}^-(x)\right]^*~\chi_{\nu,B}^{\pm}(x) = \frac{1}{2}\delta_{\nu\nu'}$ and the definition
\begin{equation}
 b_{\nu n}^{\pm\pm} := \int\limits_0^{W_2} dx\left[\chi_{\nu,B}^{\pm}(x)\right]^*~ \varphi_{n,B}^{\pm}(x)\,.
\end{equation}
Since $\varphi_{n,B}^{\pm}(x)$ vanishes for $x>W_1$, one can replace the upper limit of integration $W_2$ by $W_1$. 

Second, we project equation (\ref{eq:match}) onto modes of the narrow lead. Multiplying by $\left[\varphi_{n'}^+(x)\right]\dagg$ and integrating from $0$ to $W_1$ yields
\begin{equation}
d_{n\omega}^{++} + \sum_{\nu}d_{n\nu}^{+-}\,r_{\nu\omega} = \frac{1}{2\epsilon^2}\left(|q_n+ik_y^n|^2+\epsilon^2\right)t_{n\omega}
\label{eq:reflect2}
\end{equation}
where we have introduced the definitions (note the spinor inner product) 
\begin{equation}
 d_{n\omega}^{\pm\pm} := \int_0^{W_1} dx\left[\varphi_{n}^{\pm}(x)\right]\dagg~ \chi_{\omega}^{\pm}(x)
\end{equation}
and have again used orthogonality of the transverse wavefunctions, now in the form
\begin{equation}
\label{eq:OrthWf}
 \int_0^{W_1}dx\left[\varphi_{n}^+(x)\right]\dagg~\varphi_{n'}^+(x) =\frac{1}{2\epsilon^2} \left(|q_n+ik_y^n|^2+\epsilon^2\right)\delta_{n,n'}\,.
\end{equation}

Combining equations (\ref{eq:reflect}) and (\ref{eq:reflect2}), we obtain
\begin{equation}
\label{eq:MatEqAC}
\fl \sum_m\left(\sum_{\nu}2d_{n\nu}^{+-}b_{\nu m}^{-+} - \frac{1}{2\epsilon^2}\left(|q_n+ik_y^n|^2+\epsilon^2\right)\delta_{nm}\right)t_{m\omega} = d_{n\omega}^{+-} - d_{n\omega}^{++}
\end{equation}
which can be written as a matrix equation in the form
\begin{equation}
\sum_m M_{nm}~ t_{m\omega} = c_n\,.\
\end{equation}
This can be solved for the $t_{m\omega}$ by introducing large enough cut-offs for $m$ and $\nu$ and then inverting the now finite matrix $M$. 

The total transmission for a particle incident in mode $\omega$ from the wide side is given by
\begin{equation}
\label{eq:TWN}
 T_{\omega} = \sum_{n~ \mathrm{prop.}} T_{n\omega} = \sum_{n~ \mathrm{prop.}} \left|\frac{j_y^n}{j_y^{\omega}}\right| |t_{n\omega}|^2 
= \sum_{n~ \mathrm{prop.}} \left|\frac{k_y^n}{k_y^{\omega}}\right||t_{n\omega}|^2 \;.
\end{equation}
Finally, the conductance of the system is connected to the transmission via Landauer's formula
\begin{equation}
 G = \frac{2e^2}{h}\sum_{\omega~\mathrm{prop.}} T_{\omega}\,.
\end{equation}
In these last two equations the sums run over propagating modes only.\\



\end{document}